\documentclass[aps,prl,preprint,groupedaddress]{revtex4-1}

\usepackage{graphics}
\usepackage{graphicx}
\usepackage{amssymb}
\usepackage{amsthm}
\usepackage{natbib}

\begin{document}


\title{The Sub-bandgap Photoconductivity in InGaAs:ErAs Nanocomposites}


\author{W-D. Zhang}
\email[]{weidong.zhang@wright.edu}
\author{E. R. Brown}
\email[]{elliott.brown@wright.edu}
\affiliation{Terahertz Sensor Laboratory, Depts. of Physics and Electrical Engineering,\\ \small Wright State University, Dayton, OH, USA, 45435 }


\date{July 12, 2013}

\begin{abstract}
The photoconductions of ultrafast InGaAs:ErAs nanocomposites at low temperatures were investigated. The parabolic Tauc edge as well as the exponential Urbach tail are identified in the absorption spectrum. The Tauc edge supports that the density of states at the bottom of conduction band is proportional to the square root of energy. The Urbach edge is attributed to interband transition caused by smooth microscopic internal fields. The square root of mean-squared internal fields, whose distribution is Gaussian, is found in the order of $10^{5}$V/cm, agreeing very well with the theoretical predictions by Esser (B. ~ Esser, Phys. stat. sol. (b), vol. 51, 735 (1972)). 
\end{abstract}


\maketitle

\section{Introduction}
Ultrafast semiconductor-semimetal nanocomposite, InGaAs:ErAs, has been successfully grown with molecular beam epitaxy (MBE)\cite{Hanson}. Interestingly, the existence of ErAs within the InGaAs semiconductor crystalline lattices is in the form of self-assembled nano-scale quantum islands\cite{Sukhotin, BrownSST}. These quantum islands, with energy levels deep into the forbidden gap of InGaAs, are centers for pico-second photo-recombination. The nanocomposite has great potentials in solid-state devices such as ultrafast photoconductor, photomixer, thermoelectric converter and low-noise Schottky diode detector \cite{Sukhotin, BrownSST}. Hence, it is of importance to understand the nature of its optical and electronic properties.  Earlier the conductivity measurements revealed typical properties of non-crystalline semiconductors\cite{BrownITranNano}: Ahrennius behavior in the temperature range $\sim$195-250 K; the Mott's $\frac{1}{4}$ law in the intermediate $\sim$60-194K.
The former is dictated by electrons thermally activated from below the mobility edge to above the mobility edge; the later is consistent with the variable range hopping. In order to interpret experimental data, we found the concepts such as localized states and Mott's mobility edge were necessary.

In this paper,we report photocurrent measurements of InGaAs:ErAs nanocomposite. Similar to its electronic counterpart, the optical absorption also exhibits many features of amorphous semiconductors-Tauc edge and Urbach trail. By understanding their physical mechanisms, we are able to extract some important parameters such as density of states $N(E_{C})$ and the hopping length $L_{h}$.

\section{Experiments}

The nanocomposite was produced with the MBE method \cite{Hanson}. Along with the growth of InGaAs, Erbium and Be were deposited simultaneously.  The composition of Erbium was $0.3\%$, and the compensation from Be dopants was $5\times 10^{18}\texttt{cm}^{-3}$. Hall measurements revealed the sample was n-type; and the electron concentration was determined $n=1.2\times 10^{15}\texttt{cm}^{-3}$. 

\begin{figure}
\begin{center}
\includegraphics[scale=0.6]{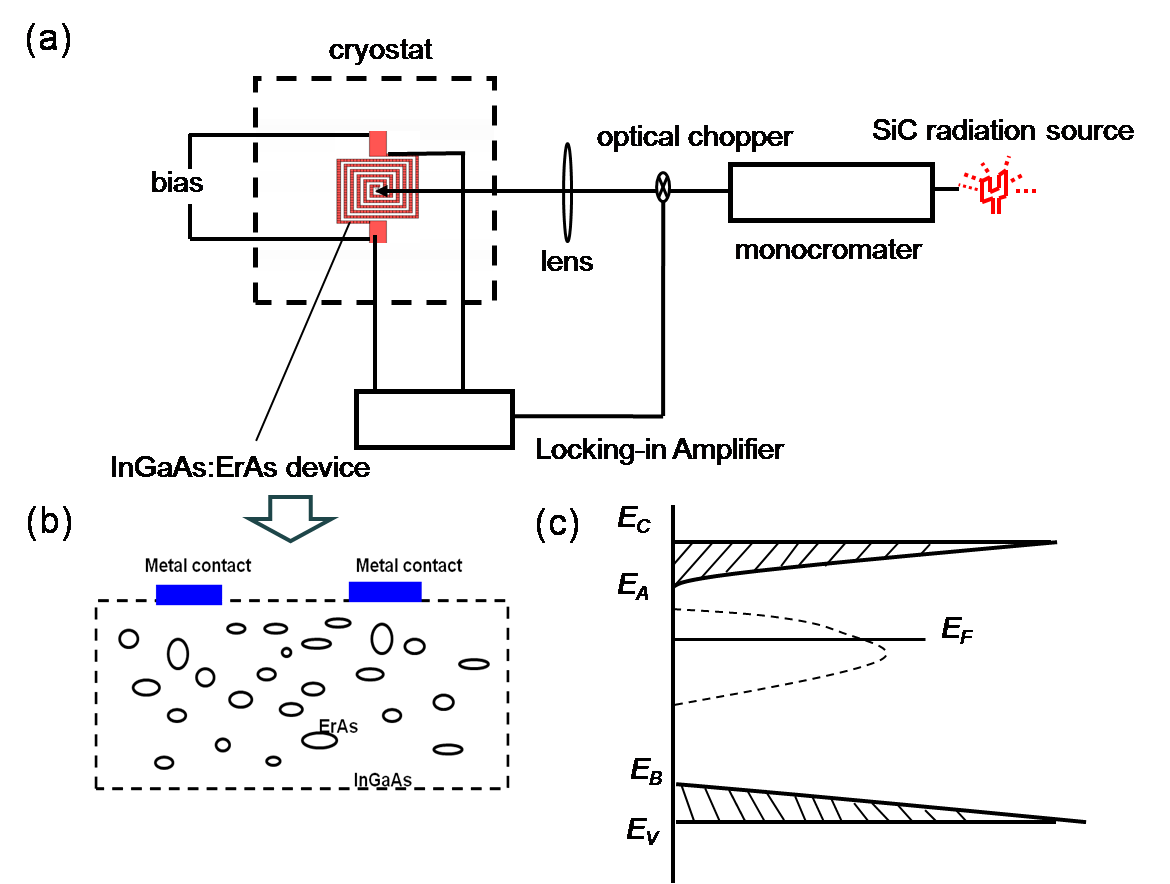}
\caption{\label{Set} (a)The experiment setup; (b) the illustration of ErAs islands in the host InGaAs lattice; and (c) the Mott mobility edges. }
\end{center}
\end{figure}

The metallic contacts on the surface of InGaAs:ErAs semiconductor was patterned into spiral squares \cite{BrownITranNano} (Fig. \ref{Set} (a) ). The distance between a pair of electrodes is $l=9\mu$m (Fig. \ref{Set} (b)). A wire-bonded diced device was mounted on the cold finger of a Gifford-McMahon close-cycle-He refrigerator. 

The illumination was a thermal source- a piece of silicon carbide filament biased at 70-100 volts. The monochromatic color was obtained with a ConerStone 260$\frac{1}{4}$m monochromator. The beam was focused with a focal lens, then passed through a Sapphire window and entered into the cryostat. Very often, there is production of "ghost" lines inside the monochromator. In order to eliminate them, two filters were placed into the beam path: one was a semi-insulating GaAs substrate and the other is a semi-insulating InP substrate with a two micron InGaAs layer on its top.

The temperature of the cold finger was measured with a DT-670B-CU silicon diode, and the readings were recorded with a Lakeshore 331 temperature controller. With the controller's heater loop, the cryostat temperature could be maintained at a targeted value.  For example, the measurements to be reported were performed at three different temperatures: 103K, 93K and 83K, respectively. 

The signal of photocurrent was measured with a lock-in amplifier.
The lock-in signal $V_{p}(\omega)$ is proportional to the absorption coefficient $\alpha(\omega)$, 
\begin{equation}
V_{p}(\omega)=\frac{V_{bias}}{l}\frac{\eta I_{0}(\omega)\alpha(\omega)}{\hbar\omega}eAR_{d}\tau\mu_{D}(1-R) \label{eq:Vp}
\end{equation}
where $A$ is the illumination area, $I_{0}(\omega)$ is the optical intensity, $\eta$ is the quantum efficiency of photogeneration, $R$ is the reflectivity, $R_{d}$ is the dc resistance, $V_{bias}$ is the bias voltage, $\tau$ is the mean lifetime of electron-hole pair, and $\mu_{D}$ is the effective electron's mobility. The optical intensity, $I_{0}(\omega)$, on the other hand, was measured with a Golay Cell detector, and then converted to a voltage signal. Hence the ratio $V_{p}(\omega)/I_{0}(\omega)$ yields all the information about $\alpha(\omega)$.
 
\section{Results and Analysis}

\begin{figure}
\begin{center}
\includegraphics[scale=0.70]{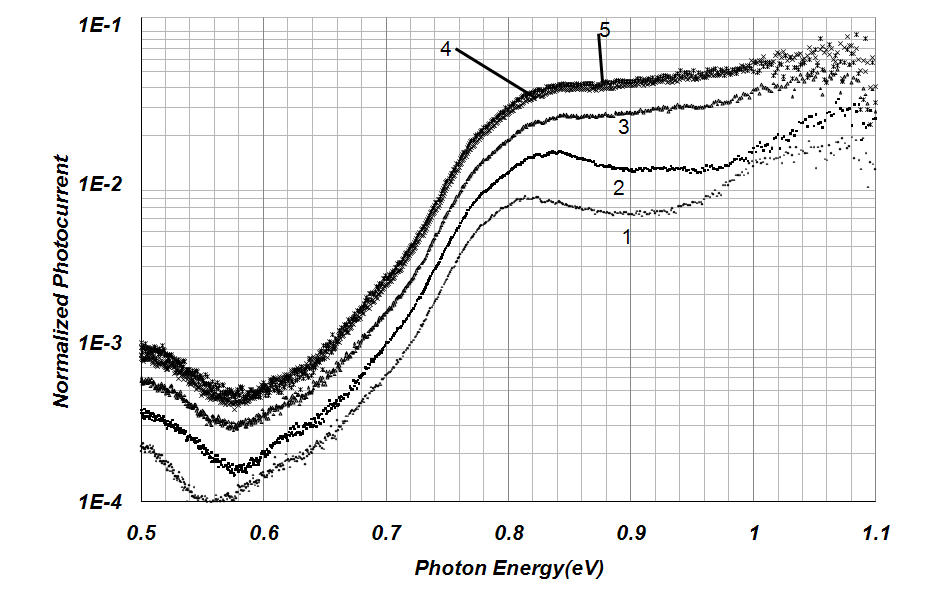}
\caption{\label{Spectra} The photocurrent at 103K. The vertical axis is $V_{p}/I_{0}$. Curve 1, $V_{bias}=1.38$; 2, $V_{bias}=2.17$; 3, $V_{bias}=3.43$; 4, $V_{bias}=5.26$; 5, $V_{bias}=6.43$ volt. }
\end{center}
\end{figure}

\begin{figure}
\begin{center}
\includegraphics[scale=0.70]{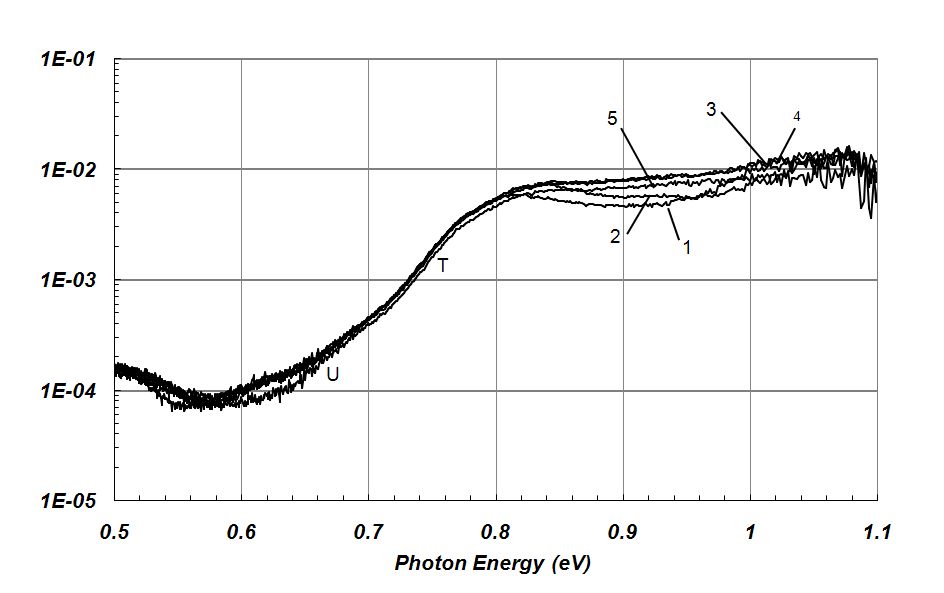}
\caption{\label{DBias} The spectrum. The vertical axis is  $V_{p}/I_{0}V_{bias}$. Curve 1, $V_{bias}=1.38$; 2, $V_{bias}=2.17$; 3, $V_{bias}=3.43$; 4, $V_{bias}=5.26$; 5, $V_{bias}=6.43$ volt.}
\end{center}
\end{figure}

The normalized photocurrents at different bias voltages $V_{bias}$ at 103K are plotted in Fig. \ref{Spectra}. The voltage-varied photoconductive measurements were also conducted at 93K and 83 K, respectively, and similar results were obtained.
When divided by $V_{bias}$, the spectral curves are mostly overlapped (Fig.\ref{DBias} ).
Near the absorption edge,there are two important regimes labeled as T, U, respectively \cite{Tauc}.

\subsection{Tauc edge}
The Tauc edge, T, is associated with the interband absorption across the bandgap \cite{Cody, Grein}. If the density of states near the conduction band edge $E_{c}$ as well as the valence band edge $E_{B}$ are proportional to the square root of energy, i.e. $N(E)=C_{a}(E-E_{A})^{1/2}$, $N(E)=C_{b}(E-E_{B})^{1/2}$, where $E_{A}$ and $E_{B}$ are mobility edges defined in Fig.\ref{Set}(c), and $C_{a}$, $C_{b}$ are parameters, the absorption coefficient is found to be parabolic \cite{Mott}
\begin{equation}
\alpha(\omega)\propto B\frac{(\hbar\omega-E_{0}^{(T)})^{2}}{\hbar\omega} \label{eq:abt}
\end{equation}
where $E_{0}^{(T)}=E_{A}-E_{B}$. 
\begin{figure}
\begin{center}
\includegraphics[scale=0.70]{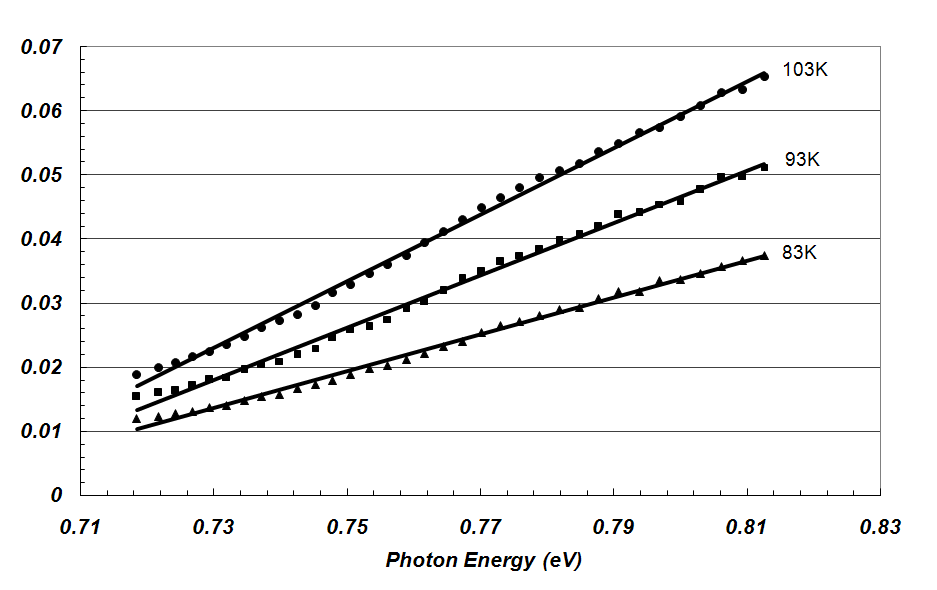}
\caption{\label{TaucEdge}The Tauc edges at 103K, 93K and 83K. The vertical axis is $\sqrt{V_{p}\times(\hbar\omega)^{2}/I_{0}}$. The fitting equations are y=0.5198x-0.3564, y=0.4071x-0.2792, y=0.288x-0.1966, respectively.}
\end{center}
\end{figure}
\begin{figure}
\begin{center}
\includegraphics[scale=0.60]{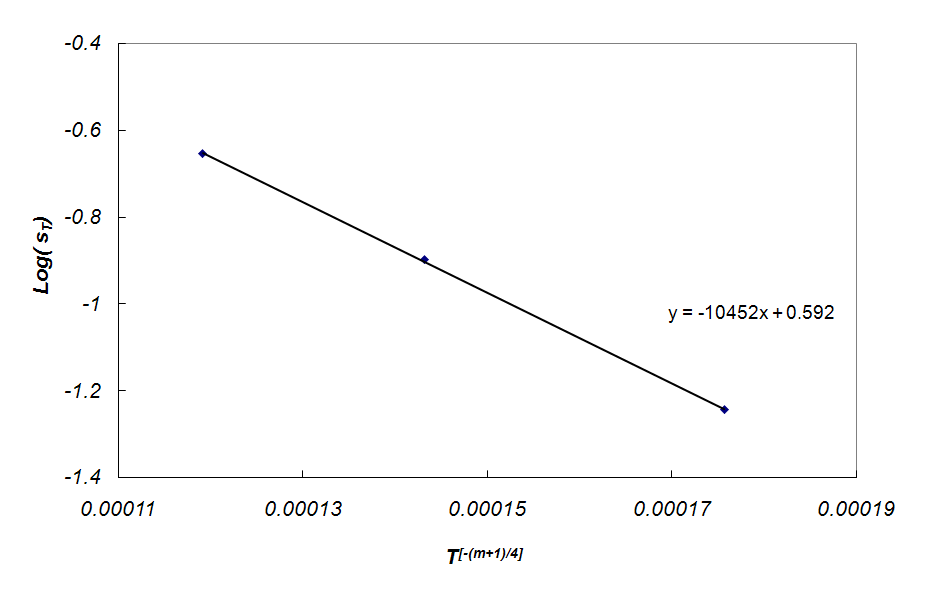}
\caption{\label{B1} The fitting of $s_{T}$.}
\end{center}
\end{figure}

The fitting results with Eq. (\ref{eq:abt}) are plotted in Fig.  \ref{TaucEdge}. The optical bandgaps $E_{0}^{(T)}$ 
are extracted: $E_{0}^{(T)}=0.686$eV at 103K; $E_{0}^{(T)}=0.686$eV at 93K; $E_{0}^{(T)}=0.683$eV at 83K, respectively, which is less than the bandgap of InGaAs - 0.8 eV. Hence, there must be localized states below the mobility edge $E_{C}$ and above $E_{A}$ ( Fig.\ref{Set}(c) ). The distance, $E_{C}-E_{A}$, is roughly (0.8-0.69)/2=0.055eV. Here we make the assumption $E_{C}-E_{A}\approx E_{B}-E_{V}$\cite{Mott}. 

In addition, we evaluated $E_{C}-E_{F}$, 0.173eV, from the conductivity-temperature curve by fitting it with
$
\sigma=\sigma_{\texttt{min}}\exp{\left(-\frac{E_{C}-E_{F}}{k_{B}T}\right)} 
$ \cite{BrownITranNano}. Knowing the position of Fermi level, we are able to calculate the free electron density from
\begin{equation}
n=\int_{E_{A}}^{\infty}N(E)\exp{\left(-\frac{E-E_{F}}{k_{B}T}\right)}dE
=C_{a}(k_{B}T)^{3/2}\sqrt{\pi}\exp{\left(-\frac{E_{A}-E_{F}}{k_{B}T}\right)}
\end{equation}
which was also measured directly from the Hall measurements. Then $C_{a}$ is estimated $1.55\times10^{19}\texttt{cm}^{-3}\texttt{eV}^{-\frac{3}{2}}$; the density of states at $E_{C}$ is deduced $N(E_{C})=3.62\times10^{18}\texttt{cm}^{-3}\texttt{eV}^{-1}$. 

Furthermore, the coefficient $B$ in Eq. (\ref{eq:abt}) can be estimated from $B=4\pi\sigma_{min}/n_{r}c(E_{C}-E_{A})$ \cite{Mott}. By taking the refractive index $n_{r}=3.5$  and $\sigma_{min}$=325 $(\Omega-\texttt{cm})^{-1}$ from the fitting of the conductivity-temperature curve, we estimate $B$ is $6.8\times10^{5}\texttt{cm}^{-1}\texttt{eV}^{-1}$. This value is in the same order as those in many amorphous semiconductors \cite{Mott}. The absorption coefficient at the band gap 0.8eV is calculated $1\times10^{4}\texttt{cm}^{-1}$, which is close to $8\times10^{3}\texttt{cm}^{-1}$ of a pure InGaAs at 100K. 

From Eq.(\ref{eq:Vp}) and Eq. (\ref{eq:abt}), the slope of $\sqrt{V_{p}(\hbar\omega)^{2}/I_{0}}$ in Fig. \ref{TaucEdge}, $s_{T}$, is
\begin{equation}
s_{T}=\lbrace\frac{V_{bias}}{l}\eta eAR_{d}\tau\mu_{D}(1-R)\rbrace^{1/2}
\end{equation}
It has a temperature dependence that
may be traced to the mobility, which is
\begin{equation}
\mu_{D}\propto \exp{(-B_{1}T^{-(m+1)/4})}
\end{equation}
where $m=(E_{C}-E_{A})/k_{B}T$, and $B_{1}$ is \cite{Mott}
\begin{equation}
B_{1}=3(\frac{8}{9})^{\frac{3}{4}}
\lbrace\frac{(E_{C}-E_{A})^{m}}{\pi N(E_{C})m^{m}k_{B}^{m+1}L_{h}^{3}}\rbrace^{1/4}
\end{equation}
where $L_{h}$ is the hopping length. 

At 103K, $m \approx 6.185$; the linear fitting of $s_{T}$ yields $B_{1}\approx10452\texttt{K}^{1.8}$ (Fig.\ref{B1}). 
With $m$, $B_{1}$, $N(E_{C})$ and $E_{C}-E_{A}$ known, $L_{h}$ is estimated to be 23.7 nm. 

\subsection{Urbach tail}

\begin{figure}
\begin{center}
\includegraphics[scale=0.70]{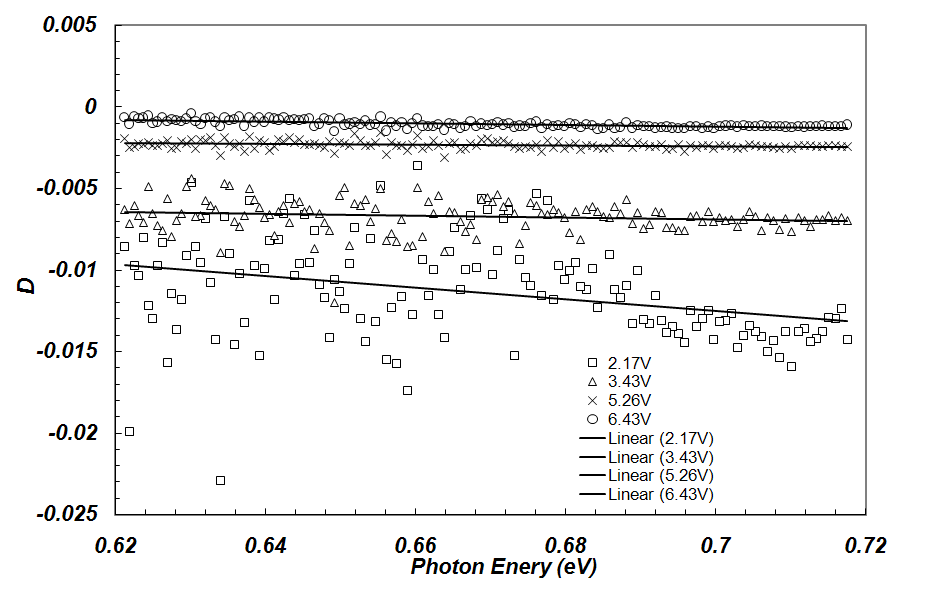}
\caption{\label{DF} The fittings of Eq.(\ref{eq:dow}). The fitting equations for bias from low to high are y=-0.0353+0.0122,y=-0.0056-0.003, y=-0.0023x-0.0008 and y=-0.0056x+0.0027, respectively. }
\end{center}
\end{figure}

The regime U over 0.62-0.72 eV is the Urbach tail.  There are a few available theories about the origin of exponential Urbach edge. For example, a popular one is the broadening of excitonic lines in strong microscopic (internal) electric fields, (the so-called Dow-Redfield's theory \cite{Mott},\cite{Dow}). If the Urbach edge is due to Dow-Redfield absorption, then the variations of absorption coefficients in response to the external fields, $F_{l}=V_{a}/l$, should obey \cite{Mott}
\begin{equation}
\frac{\Delta \alpha}{\alpha}=\frac{\alpha|_{V_{bias}}-\alpha|_{V_{bias}=0}}{\alpha}\propto(E_{g}-W-\hbar\omega)^{2/3}F_{l}^{2}
\end{equation}
where $E_{g}$ is the optical bandgap, and $W$ is the binding energy of exciton. However, the absorption curves plotted in Fig. \ref{DBias} are almost the same at different external fields. Furthermore, 
the equation above predicts
\begin{equation} D=-\left[ \frac{(\alpha-\alpha_{1})}{\alpha_{1}V_{bias}^{2}} \right] ^{3/2}\propto E_{g}-W-\hbar\omega \label{eq:dow}
\end{equation}
where $\alpha_{1}=\alpha|_{V_{bias}=1.38V}$ is approximated as $\alpha|_{V_{bias}=0}$. The fittings in Fig. \ref{DF} show significant inconsistencies among the extracted $(E_{g}-W)$: 0.35eV at $V_{bias}=2.17$V; -0.54 eV at $V_{bias}=3.43$V; -0.35eV at $V_{bias}=5.26$V ; 0.48 eV at $V_{bias}=6.43$V, respectively. Hence the excitonic transition is not the primary physical reason behind the Urbach edge.
 
\begin{figure}
\begin{center}
\includegraphics[scale=0.70]{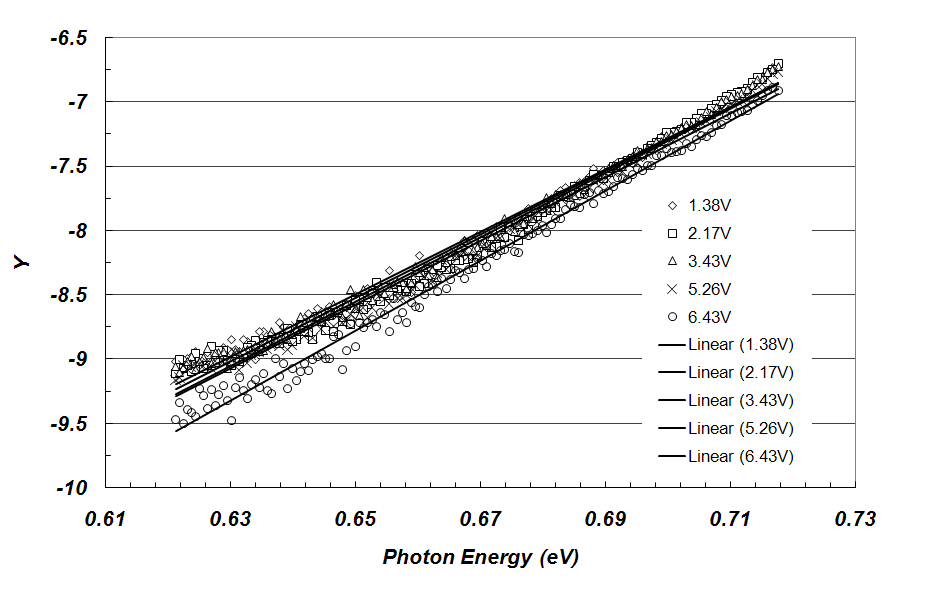}
\caption{\label{Urbachvoltage103}The Urbach edges at 103K.
The fitting equations for bias from low to high are y =24.251x-24.257, y=25.015x-24.809, y=24.637x-24.537,  y=24.755x-24.663, y=27.123x-26.406 respectively.}
\end{center}
\end{figure}
\begin{figure}
\begin{center}
\includegraphics[scale=0.70]{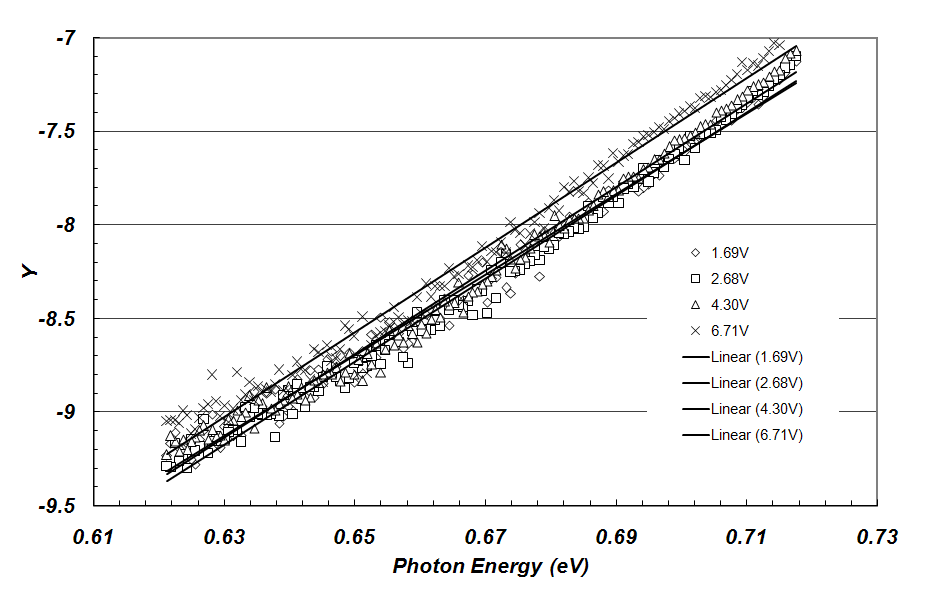}
\caption{\label{Urbachvoltage93}The Urbach edges at 93K.
The fitting equations for bias from low to high bias are y=21.551x-22.706,y=22.158x-23.134, y=22.302x-23.186, y=22.628x-23.280, respectively.}
\end{center}
\end{figure}
\begin{figure}
\begin{center}
\includegraphics[scale=0.70]{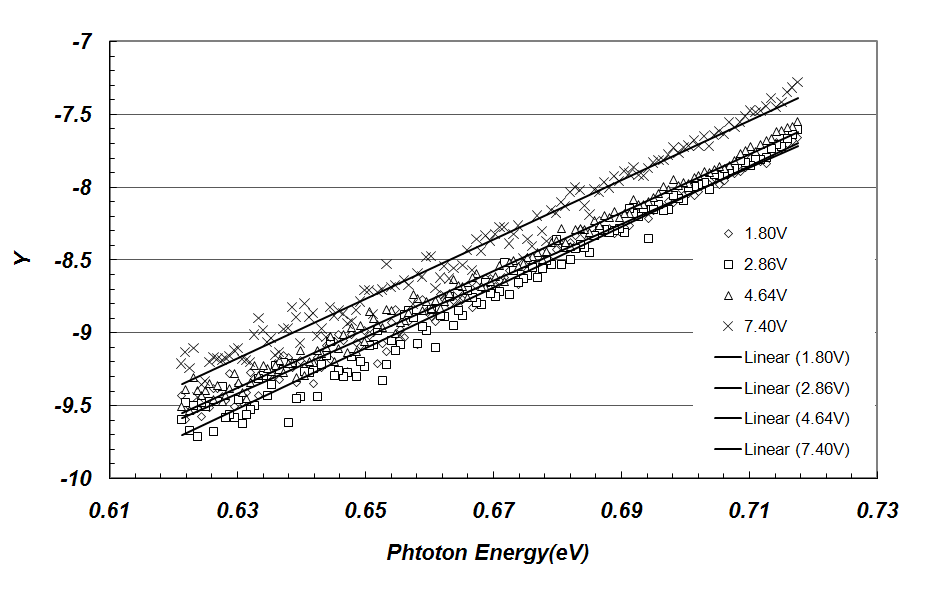}
\caption{\label{Urbachvoltage83}The Urbach edges at 83K.
The fitting equations for bias from low to high are y=19.414x-21.648, y=20.830x-22.644, y=20.004x-21.975, y=20.347x-21.992, respectively.}
\end{center}
\end{figure}
Because the Urbach edge has very weak external field dependence, we apply a smooth random field theory given by Esser as the possible explanation of its origin \cite{Esser}. The critical assumptions are: microscopic internal fields are random,continuous and smooth; their functional distribution falls into Gaussian. At low temperature, the absorption coefficient is given by
\begin{equation}
\alpha(\omega)=A_{1}\frac{\sqrt{E_{g}-\hbar\omega}}%
{\hbar \omega}[1+\frac{e^{2}F_{l}^{2}}{3\psi_{2}}\Gamma(E_{g}-\hbar\omega)]e^{\Gamma(\hbar\omega-E_{g})} \label{eq:Uedge}
\end{equation}
where $A_{1}$ is a constant. 
In particular, the exponential parameter $\Gamma$ is related to $\psi_{2}$-the mean square value of internal fields by
\begin{equation}
\Gamma=\left(\frac{\hbar^{2}e^{2}\psi_{2}}{36\mu}\right)^{-1/3}
\end{equation}
where $\mu$ is reduced effective mass of electron-hole pair. Thus, $\Gamma$ could be extracted from the fitting of
\begin{equation}
Y=\ln{\left[\frac{V_{p}(\omega)}{I_{0}(\omega)}(\hbar\omega)^{2}\right]}-0.5\ln{(E_{g}-\hbar\omega)}-
\ln{\left[1+\frac{e^{2}F_{l}^{2}\Gamma(E_{g}-\hbar\omega)}{3\psi_{2}}\right]}\propto \Gamma(\hbar\omega-E_{g}) \label{eq:Uedgefitting}
\end{equation}
The equation is further simplified as the external field is less than the internal fields, $F_{l}<\sqrt{\psi_{2}}$, thus the the third term of $Y$ is approximated as zero, 
\begin{equation}
Y\approx\ln{\left[\frac{V_{p}(\omega)}{I_{0}(\omega)}(\hbar\omega)^{2}\right]}-0.5\ln{(E_{g}-\hbar\omega)}
\propto \Gamma(\hbar\omega-E_{g}) \label{eq:UedgefY}
\end{equation}

Now the $Y$ values can be computed from Eq. (\ref{eq:UedgefY}). Here $V_{p}(\omega)/I(\omega)$ can be found from Fig. \ref{DBias}; $E_{g}$ is taken as InGaAs's 0.8 eV; $\mu$ is approximately taken as the effective electron mass $0.04m_{0}$ of InGaAs. The fittings of $Y$ at 103K (Fig. \ref{Urbachvoltage103}), 93K (Fig. \ref{Urbachvoltage93} ), and 83K (Fig. \ref{Urbachvoltage83}) agree well with Eq. (\ref{eq:Uedgefitting}) and Eq. (\ref{eq:Uedge}). Specifically, at 93K and 83K, the values of $\Gamma$ at different $V_{bias}$  are almost the same.  At 103K, the values of $\Gamma$ at the first four $V_{bias}$ are almost the same; the one at the bias 6.43V deviates the trend a little bit (about 12\%), but within the range of errors.  

\begin{table}[h]
\caption{The $\Gamma$ parameters and the square root of mean square internal fields} 
\centering 
\begin{tabular}{ c c c c  } 
\hline\hline 
  $  $  & $103K $ & $93K$ & $83K$  \\ [0.5ex] 
\hline 
 $\Gamma (\texttt{eV}^{-1})$ & 24.2 & 22.2 & 20.2 \\
\hline 
 $\sqrt{\psi_{2}}$ (V/cm) & $3.9\times 10^{5}$ & $ 4.4 \times 10^{5}$ & $ 5.1\times 10^{5}$ \\
\hline
\end{tabular}
\label{table:ta1} 
\end{table}

At 103K, $\Gamma\approx24.2\texttt{eV}^{-1}$, $\sqrt{\psi_{2}}$ is then estimated $\sim3.9\times 10^{5}$ V/cm; at 93K, $\Gamma\approx22.2\texttt{eV}^{-1}$, $\sqrt{\psi_{2}}$ is $\sim4.4\times 10^{5}$V/cm; and at 93K, $\Gamma\approx20.2\texttt{eV}^{-1}$, $\sqrt{\psi_{2}}$ is $\sim5.1\times 10^{5}$V/cm (Table \ref{table:ta1}). Being in the range $10^{5}-10^{6}$ V/cm,  the values of $\sqrt{\psi_{2}}$ agree with the theoretical predictions. The external fields are in the order of  $10^{3}$ V/cm, which are 2-3 order less than the square root of $\psi_{2}$. This justifies the approximation in Eq. (\ref{eq:UedgefY}). The $\Gamma$ values are close to what have already been seen (15-22 $\texttt{eV}^{-1}$ ) in a number of non-crystalline materials \cite{Mott}. 

\section{Conclusions}

In summary, the optical absorption of InGaAs:ErAs nanocomposite exhibits many features of non-crystalline structures. There is the parabolic Tauc edge close to the bandgap. Below it is the exponential Urbach edge. 
The parabolic Tauc edge allows us to conclude shallow localized states below the conduction band mobility edge $E_{c}$ and the density of states in this range is the function of square root of energy.
The dependence of Urbach tails on external fields allow us to determine its cause is not the Dow-Redfield excitonic effect, but an interband absoprtion under the influence of smooth microscopic internal fields. 
Furthermore, the square root of mean square of internal fields is evaluated in the order of $10^{5}$V/cm, agreeing well with the theory given by Esser.  Finally, some of important parameters, such as density of states $N(E_{C})$ as well as the hopping length $L_{h}$, are estimated. 
\\

\begin{acknowledgments}
This material is based upon work supported by, or in part by, the U. S. Army Research Laboratory and the U. S. Army Research Office under contract/grant number W911NF1210496.  The authors also thank Dr. H. Lu for information related to the InGaAs:ErAs samples.
\end{acknowledgments}

\bibliography{ZhangPP}

\end{document}